\documentclass{article}
\usepackage{dcase2019,amsmath,url,times,booktabs,tabularx}
\usepackage{array,multirow}
\usepackage{multicol}
\usepackage{float}
\usepackage{cite}

\usepackage[pdftex]{graphicx}

\newcolumntype{C}[1]{>{\centering\arraybackslash}p{#1}}
\newcolumntype{L}[1]{>{\raggedright\arraybackslash}p{#1}}
\newcolumntype{R}[1]{>{\raggedleft\arraybackslash}p{#1}}

\title{MIMII Dataset: Sound Dataset for\\ Malfunctioning Industrial Machine Investigation and Inspection}

\name{Harsh Purohit,
      Ryo Tanabe,
      Kenji Ichige, 
      Takashi Endo,
      }
\secondlinename{
      Yuki Nikaido,
      Kaori Suefusa,
      and Yohei Kawaguchi
      }
\address{ Research and Development Group, Hitachi, Ltd.\\
1-280, Higashi-koigakubo, Kokubunji, Tokyo 185-8601, Japan \\
\{harsh\_pramodbhai.purohit.yf, yohei.kawaguchi.xk\}@hitachi.com}

\begin{document}

\ninept
\maketitle

\begin{sloppy}

\begin{abstract}
Factory machinery is prone to failure or breakdown, resulting in significant expenses for companies. 
Hence, there is a rising interest in machine monitoring using different sensors including microphones. 
In the scientific community, the emergence of public datasets has led to advancements in acoustic detection and classification of scenes and events,
but there are no public datasets that focus on the sound of industrial machines under normal and anomalous operating conditions in real factory environments.
In this paper, we present a new dataset of industrial machine sounds 
that we call a sound dataset for malfunctioning industrial machine investigation and inspection (MIMII dataset). 
Normal sounds were recorded for different types of industrial machines (i.e., valves, pumps, fans, and slide rails), 
and to resemble a real-life scenario, various anomalous sounds were recorded (e.g., contamination, leakage, rotating unbalance, and rail damage).
The purpose of releasing the MIMII dataset is to assist the machine-learning and signal-processing community with their development of automated facility maintenance.
\end{abstract}


\section{Introduction}
\label{sec:intro}

The increasing demand for automatic machine inspection stems from the need for a better quality of factory equipment maintenance. 
The discovery of malfunctioning machine parts mainly depends on the experience of the field engineer, 
but currently there is a shortage of field experts due to the increased number of requests for inspection.
An efficient and affordable solution to this problem is urgently required.

In the past decade, industrial Internet of Things (IoT) and data-driven techniques have been revolutionizing the manufacturing industry, 
and different approaches have been undertaken for monitoring the state of machinery. 
Examples include vibration sensor-based approaches \cite{yu2013model, ishibashi2019modelling, carden2004vibration, galloway2016diagnosis}, 
temperature sensor-based approaches \cite{Lodewijks}, and pressure sensor-based approaches \cite{Salikhov}. 
Another approach is to detect anomalies from sound by using technologies for acoustic scene classification and event detection 
\cite{koizumi2019sniper, kawachi2019two, yamaguchi2019adaflow, kawaguchi2019anomaly, koizumi2018unsupervised, kawaguchi2017can, kawaguchi2018anomaly}. 
Remarkable advancements have been made in the classification of acoustic scenes and the detection of acoustic events,
and there are many promising state-of-the-art studies in this vein \cite{mesaros2018detection, phaye2019subspectralnet, podwinska2019acoustic}.
It is clear that the emergence of numerous open benchmark datasets \cite{audioset, freesound, sins, toyadmos} is essential for the advancement of the research field. 
However, to the best of our knowledge, there is no public dataset that contains different types of machine sounds in real factory environments. 

In this paper, we introduce a new dataset of machine sounds under normal and anomalous operating conditions in real factory environments. 
We include the sound of four machine types---(i) valves, (ii) pumps, (iii) fans, and (iv) slide rails---and
for each type of machine, we consider seven different product models. 
We assume that the main task is to find an anomalous condition of the machine during a 10-second sound segment in an unsupervised learning situation. 
In other words, only normal machine sounds can be used in the training phase, 
and we have to correctly distinguish between a normal machine sound and an abnormal machine sound in the test phase.
The main contributions of this paper are as follows:
(1) We created an open dataset for malfunctioning industrial machine investigation and inspection (MIMII), the first of its kind. 
We have released this dataset,
and it is freely available for download at \url{https://zenodo.org/record/3384388}.
This dataset contains 26,092 sound files for normal conditions of four different machine types. 
It also contains real-life anomalous sound files for each category of the machines. 
(2) Using our developed dataset, we have explored an autoencoder-based model for each type of machine with various noise conditions. 
These results can be taken as a benchmark to improve the accuracy of anomaly detection in the MIMII dataset.

In Section \ref{sec:2} of this paper, we describe our recording environment and the setup. 
The details of the dataset content are provided in Section \ref{sec:3}. 
The autoencoder-based detection benchmark and results are discussed in Section \ref{sec:4}. 
We conclude in Section \ref{sec:5} with a brief summary and mention of future work.

\section{Recording Environment and Setup}
\label{sec:2}

\begin{figure}[t]
\begin{center}
\includegraphics[width=0.6\hsize,clip]{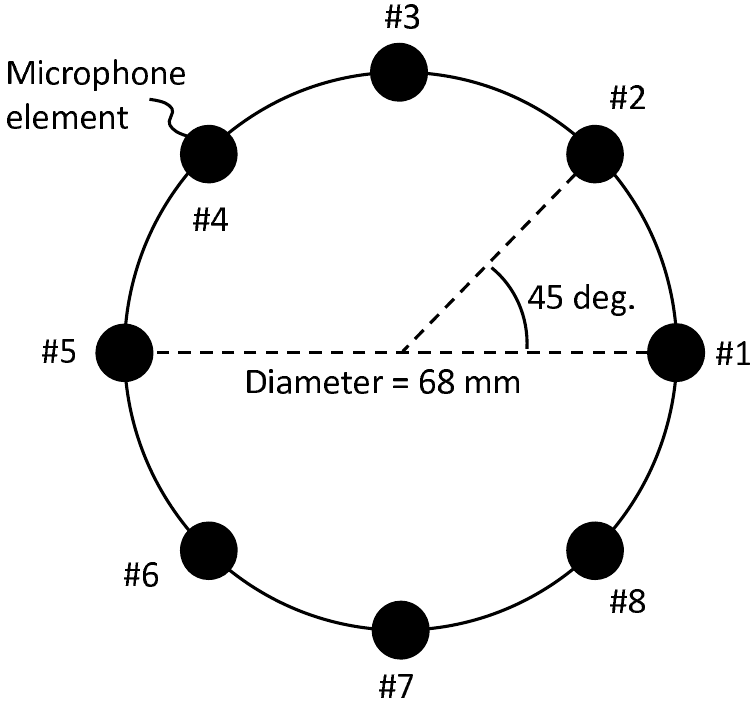}
\caption{Circular microphone array.}
\label{fig:mic}
\end{center}
\end{figure}
\begin{figure}[t]
\begin{center}
\includegraphics[width=0.95\hsize,clip]{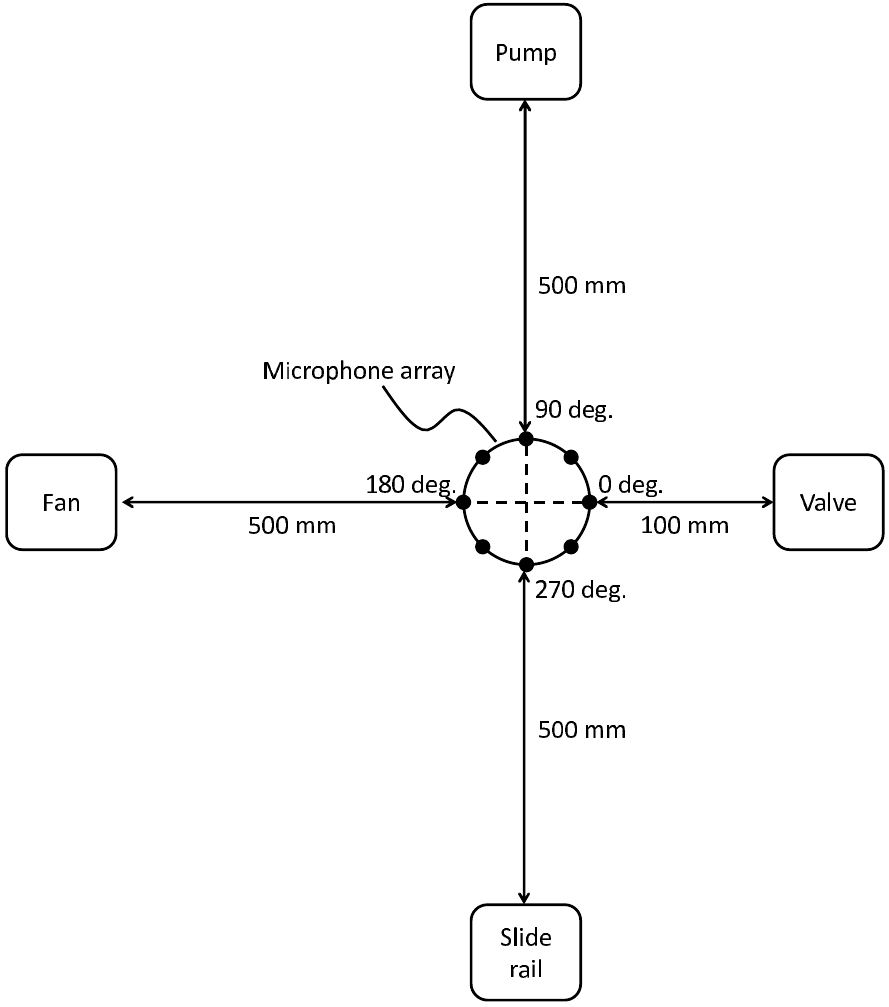}
\caption{Schematic experimental setup for dataset recording.}
\label{fig:setup}
\end{center}
\end{figure}
The dataset was collected using a TAMAGO-03 microphone manufactured by \textit{System In Frontier Inc.} \cite{tamago}. 
It is a circular microphone array that consists of eight distinct microphones, the details of which are shown in Fig. \ref{fig:mic}. 
By using this microphone array, we can evaluate not only single-channel-based approaches but also multi-channel-based ones.
The microphone array was kept at a distance of 50 cm from the machine (10 cm in the case of valves), 
and 10-second sound segments were recorded. 
The dataset contains eight separate channels for each segment. 
Figure \ref{fig:setup} depicts the recording setup with the direction and distance for each kind of machine. 
Note that each machine sound was recorded in a separate session. 
Under the running condition, the sound of the machine was recorded as 16-bit audio signals sampled at 16 kHz in a reverberant environment. 
Apart from the target machine sound, 
background noise in multiple real factories was continuously recorded and later mixed with the target machine sound for simulating real environments. 
For recording the background noise, we used the same microphone array as for the target machine sound.

\section{Dataset Content}
\label{sec:3}
The MIMII dataset contains the sound of four different types of machines: valves, pumps, fans, and slide rails. 
The valves are solenoid valves that are repeatedly opened and closed.
The pumps are water pumps that drain water from a pool and discharge water to the pool continuously.
The fans represent industrial fans, which are used to provide a continuous flow of gas or air in factories.
The slide rails in this paper represent linear slide systems, which consist of a moving platform and a stage base.
The types of the sounds produced by the machines are stationary and non-stationary, have different features, and have different degrees of difficulty. 
Figure \ref{fig:spec} depicts a power spectrogram of the sound of all four types of machines, 
clearly showing that each machine has its unique sound characteristics. 
\begin{figure*}[t]
\begin{center}
\begin{minipage}{0.48\hsize}
\begin{center}
\includegraphics[width=0.97\hsize,clip]{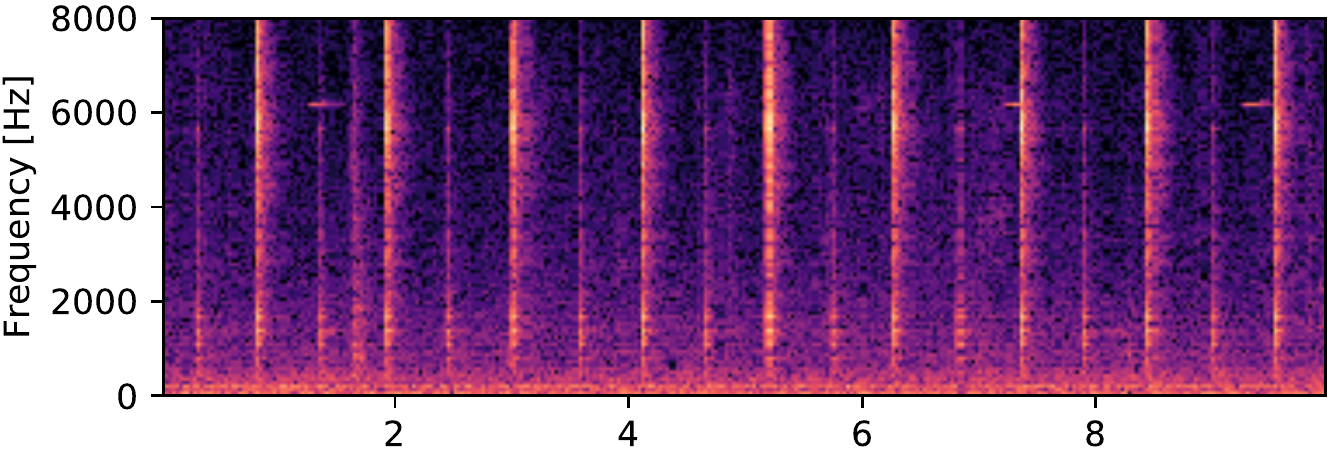}\\
(a) Valve (model ID: 00)\\
\end{center}
\end{minipage}
\begin{minipage}{0.48\hsize}
\begin{center}
\includegraphics[width=0.97\hsize,clip]{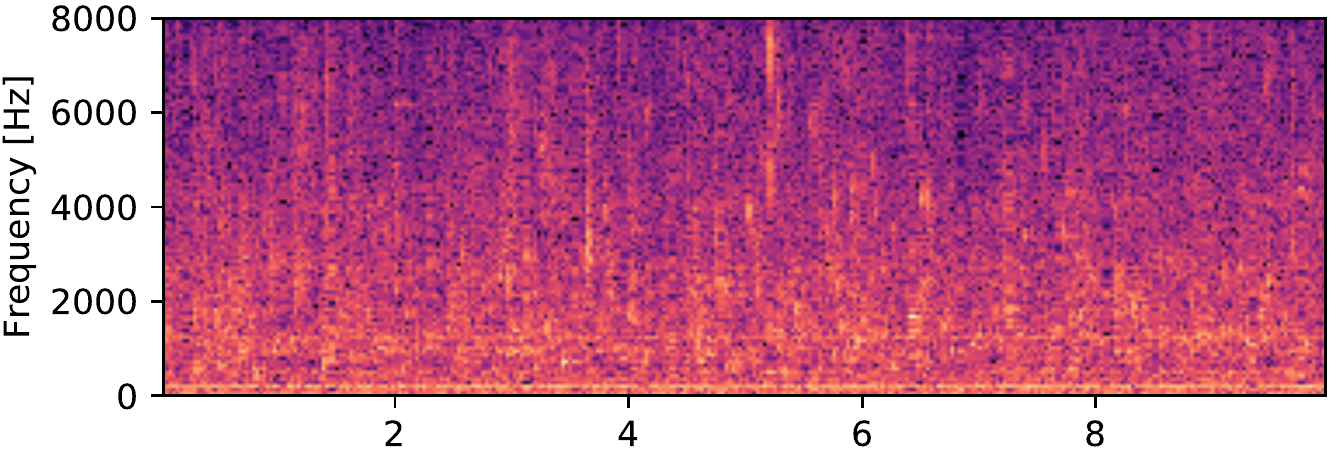}\\
(b) Pump (model ID: 00)\\
\end{center}
\end{minipage}
\begin{minipage}{0.48\hsize}
\begin{center}
\includegraphics[width=0.97\hsize,clip]{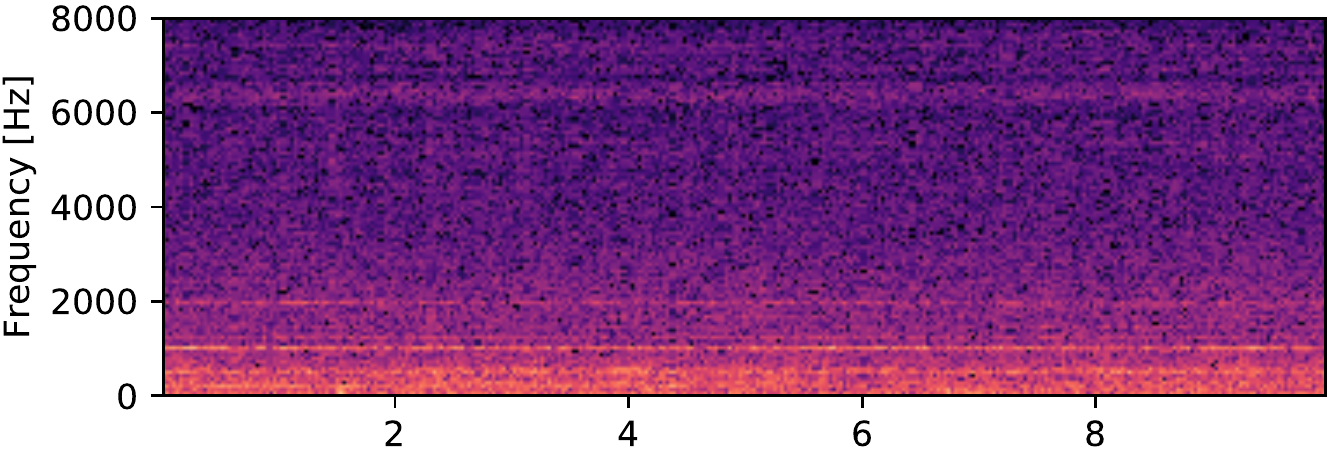}\\
(c) Fan (model ID: 00)\\
\end{center}
\end{minipage}
\begin{minipage}{0.48\hsize}
\begin{center}
\includegraphics[width=0.97\hsize,clip]{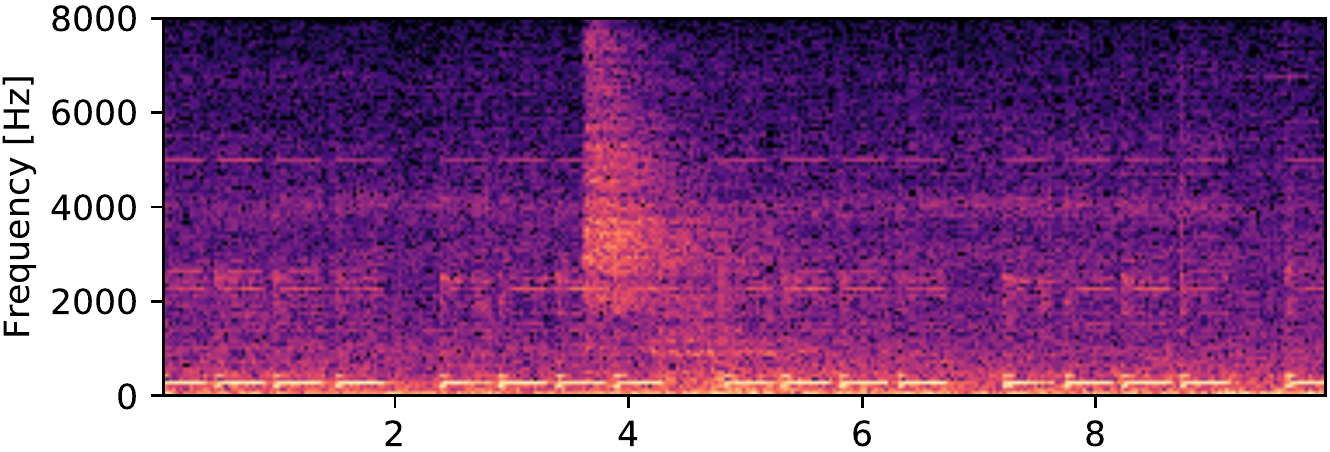}\\
(d) Slide rail (model ID: 00)\\
\end{center}
\end{minipage}
\caption{Examples of power spectrograms under normal condition at $6$-dB SNR.}
\label{fig:spec}
\end{center}
\end{figure*}

The list of sound files for each machine type is provided in Table \ref{T1}. 
Each type of machine includes seven individual machines. 
Individual machines may be of a different product model. 
We know that large datasets incorporating real-life complexity are needed to effectively train the models,
so we recorded a total of 26,092 normal sound segments for all individual machines.
In addition to this, different real-life anomalous scenarios have been considered for each kind of machine:
contamination, leakage, rotating unbalance, rail damage, etc.
The various running conditions are listed in Table \ref{T2}. 
The number of sound segments for each anomalous sound for each different type of machine is small 
because we regard the main target of our dataset as an unsupervised learning scenario and regard the anomalous segments as a part of the test data.

\begin{table}
\begin{center}
\caption{MIMII dataset content details.}
\label{T1}
\begin{tabular}{l l r r}
\hline
\multicolumn{2}{c}{\begin{tabular}{@{\hskip0pt}l@{\hskip0pt}} \textbf{Machine type / }\\ \textbf{model ID} \end{tabular}} &
\begin{tabular}{@{\hskip0pt}r@{\hskip0pt}} \textbf{Segments}\\ \textbf{for normal}\\ \textbf{condition} \end{tabular} & 
\begin{tabular}{@{\hskip0pt}r@{\hskip0pt}} \textbf{Segments}\\ \textbf{for anomalous}\\ \textbf{condition} \end{tabular} \\
\hline
\multirow{7}{*}{Valve}
& 00 & $991$ & $119$ \\
& 01 & $869$ & $120$ \\
& 02 & $708$ & $120$ \\
& 03 & $963$ & $120$ \\
& 04 & $1000$ & $120$ \\
& 05 & $999$ & $400$ \\ 
& 06 & $992$ & $120$ \\ \hline
\multirow{7}{*}{Pump}
& 00 & $1006$ & $143$ \\
& 01 & $1003$ & $116$ \\
& 02 & $1005$ & $111$ \\ 
& 03 & $706$ & $113$ \\ 
& 04 & $702$ & $100$ \\ 
& 05 & $1008$ & $248$ \\
& 06 & $1036$ & $102$ \\ \hline
\multirow{7}{*}{Fan}
& 00 & $1011$ & $407$ \\
& 01 & $1034$ & $407$ \\
& 02 & $1016$ & $359$ \\
& 03 & $1012$ & $358$ \\
& 04 & $1033$ & $348$ \\
& 05 & $1109$ & $349$ \\
& 06 & $1015$ & $361$ \\ \hline
\multirow{7}{*}{Slide rail}
& 00 & $1068$ & $356$ \\
& 01 & $1068$ & $178$ \\
& 02 & $1068$ & $267$ \\
& 03 & $1068$ & $178$ \\
& 04 & $534$ & $178$ \\
& 05 & $534$ & $178$ \\
& 06 & $534$ & $89$ \\ \hline
\multicolumn{2}{l}{Total} & $26092$ & $6065$ \\ \hline
\end{tabular}\\
\end{center}
\end{table}
\begin{table}
\begin{center}
\caption{List of operations and anomalous conditions.}
\label{T2}
\begin{tabular}{l c c}
\hline
\begin{tabular}{@{\hskip0pt}l@{\hskip0pt}} \textbf{Machine}\\ \textbf{type} \end{tabular} & 
\begin{tabular}{c} \textbf{Operations} \end{tabular} & 
\begin{tabular}{c} \textbf{Examples of}\\ \textbf{anomalous}\\ \textbf{conditions} \end{tabular} \\
\hline
Valve & 
\begin{tabular}{c} Open / close repeat\\ with different timing \end{tabular} &
\begin{tabular}{c} More than\\ two kinds of\\ contamination \end{tabular} \\
\hline
Pump & 
\begin{tabular}{c} Suction from /\\ discharge to\\ a water pool \end{tabular} &
\begin{tabular}{c} Leakage,\\ contamination,\\ clogging, etc. \end{tabular} \\
\hline
Fan & 
\begin{tabular}{c} Normal operation \end{tabular} &
\begin{tabular}{c} Unbalanced,\\ voltage change,\\ clogging, etc. \end{tabular} \\
\hline
Slide rail & 
\begin{tabular}{c} Slide repeat at\\ different speeds \end{tabular} &
\begin{tabular}{c} Rail damage,\\ loose belt,\\ no grease, etc. \end{tabular} \\
\hline
\end{tabular}
\end{center}
\end{table}

As explained in Section \ref{sec:2}, the background noise recorded in multiple real factories was mixed with the target machine sound. 
Eight channels are considered separately when mixing the original sounds with the noise. 
For a certain signal-to-noise ratio (SNR) $\gamma$ dB, the noise-mixed data of each machine model were created by the following steps:
\begin{enumerate}
\item The average power over all segments of the machine models, $a$, was calculated.
\item For each segment $i$ from the machine model,
\begin{enumerate}
\item a background-noise segment $j$ is randomly selected, and its power $b_j$ is tuned so that $\gamma = 10 \log_{10}\left( a /b_j \right)$; and
\item the noise-mixed data is calculated by adding the target-machine segment $i$ and the power-tuned background-noise segment $j$.\
\end{enumerate}
\end{enumerate}

\section{Experiment}
\label{sec:4}

An example of benchmarking is presented in this section.
Our main goal is to detect anomalous sounds in an unsupervised learning scenario, as discussed in Section \ref{sec:intro}.
Several studies have successfully used autoencoders for unsupervised anomaly detection 
\cite{tagawa2015structured, marchi2015novel, kawaguchi2017can, oh2018residual}, 
so here, we evaluate an autoencoder-based unsupervised anomaly detector.

We used only the first channel of microphones (``No. 1'' in Fig. \ref{fig:mic}).
We consider log-Mel spectrogram as an input feature. 
To calculate the Mel spectrogram, we consider a frame size of 1024, a hop size of 512, and 64 mel filters in this experiment. 
Five frames have been combined to initiate our 320 dimensional input feature vector $\mathbf{x}$. 
The parameters of the encoder and decoder neural networks (i.e., $\theta = (\theta_{e},\theta_{d})$) are trained to minimize the loss function given as
\begin{align}
L_{AE} (\theta_{e},\theta_{d}) = \left\lVert \mathbf{x}- D( E(\mathbf{x} ~| ~\theta_{e})~| ~\theta_{d})\right\rVert_2^2.
\label{re}
\end{align}
Our basic assumption is that this trained model will have a high reconstruction error for anomalous machine sounds. 
The autoencoder network structure for the experiment is summarized as follows. 
The encoder network ($E(\cdot)$) comprises $FC(Input, 64, ReLU)$; $FC(64, 64, ReLU)$; and $FC(64, 8, ReLU)$, 
and the decoder network ($D(\cdot)$) incorporates $FC(8, 64, ReLU)$; $FC(64, 64, ReLU)$; and $FC(64, Output, none)$, 
where $FC(a, b, f)$ means a fully connected layer with $a$ input neurons, $b$ output neurons, and activation function $f$.
The ReLUs are Rectified Linear Units \cite{jarrett2009best}. 
The network is trained by the Adam \cite{kingma2014adam} optimization technique for 50 epochs. 

For each machine type and model ID, all the segments were split into a training dataset and a test dataset.
All the anomalous segments were regarded as the test dataset,
the same number of normal segments was randomly selected and regarded as the test dataset,
and all the rest of the normal segments were regarded as the training dataset.
By using the training dataset consisting only of normal ones, different autoencoders were trained for each machine type and model ID.
Anomaly detection was performed for each segment by thresholding the reconstruction error averaged over ten seconds,
and the area under the curve (AUC) values were calculated for the test dataset for each machine type and model ID.
In addition to this, we considered different levels of SNR (with factory noise): 
for example, $6$ dB, $0$ dB, and $-6$ dB. 

Table \ref{T3} lists the AUCs averaged over three training runs with independent initializations.
It is clear here that the AUCs for valves are lower than the other machines.
Sound signals of valves are non-stationary---in particular, impulsive and sparse in time---and the reconstruction error averaged over time tends to be small.
That makes it difficult to detect anomalies for valves.
In contrast, it is easier to detect anomalies for fans, as the sound signals of fans are stationary.
Moreover, for some machine models, the AUC decreases rapidly as the noise level increases. 
These results indicate that we need to solve the degradation caused by non-stationarity and noise for unsupervised anomalous sound detection. 

\begin{table}
\begin{center}
\caption{AUCs for all machines.}
\label{T3}
\begin{tabular}{l l r r r}
\hline
\multicolumn{2}{c}{\multirow{2}{*}{\begin{tabular}{@{\hskip0pt}l@{\hskip0pt}} \textbf{Machine type / }\\ \textbf{model ID} \end{tabular}}} &
\multicolumn{3}{c}{\textbf{Input SNR}} \\ \cline{3-5}
 & & \textbf{$6$ dB} & \textbf{$0$ dB} & \textbf{$-6$ dB} \\
\hline
\multirow{8}{*}{Valve}
& 00 & $0.68$ & $0.55$ & $0.62$ \\
& 01 & $0.77$ & $0.71$ & $0.61$ \\
& 02 & $0.66$ & $0.59$ & $0.57$ \\
& 03 & $0.70$ & $0.65$ & $0.44$ \\
& 04 & $0.64$ & $0.65$ & $0.50$ \\
& 05 & $0.52$ & $0.48$ & $0.44$ \\
& 06 & $0.70$ & $0.66$ & $0.53$ \\ \cline{2-5}
& Avg. & $0.67$ & $0.61$ & $0.53$ \\ \hline
\multirow{8}{*}{Pump}
& 00 & $0.84$ & $0.65$ & $0.58$ \\
& 01 & $0.98$ & $0.90$ & $0.73$ \\
& 02 & $0.45$ & $0.46$ & $0.52$ \\
& 03 & $0.79$ & $0.81$ & $0.75$ \\
& 04 & $0.99$ & $0.95$ & $0.93$ \\
& 05 & $0.66$ & $0.66$ & $0.64$ \\
& 06 & $0.94$ & $0.76$ & $0.61$ \\ \cline{2-5}
& Avg. & $0.81$ & $0.74$ & $0.68$ \\ \hline
\multirow{8}{*}{Fan}
& 00 & $0.75$ & $0.63$ & $0.57$ \\
& 01 & $0.97$ & $0.90$ & $0.70$ \\
& 02 & $0.99$ & $0.83$ & $0.68$ \\
& 03 & $1.00$ & $0.89$ & $0.70$ \\
& 04 & $0.92$ & $0.75$ & $0.57$ \\
& 05 & $0.95$ & $0.90$ & $0.83$ \\
& 06 & $0.99$ & $0.97$ & $0.83$ \\ \cline{2-5}
& Avg. & $0.94$ & $0.84$ & $0.70$ \\ \hline
\multirow{8}{*}{Slide rail}
& 00 & $0.99$ & $0.99$ & $0.93$ \\
& 01 & $0.94$ & $0.90$ & $0.83$ \\
& 02 & $0.93$ & $0.79$ & $0.74$ \\
& 03 & $0.99$ & $0.85$ & $0.71$ \\
& 04 & $0.88$ & $0.78$ & $0.61$ \\
& 05 & $0.84$ & $0.70$ & $0.60$ \\
& 06 & $0.71$ & $0.56$ & $0.52$ \\ \cline{2-5}
& Avg. & $0.90$ & $0.80$ & $0.70$ \\ \hline
\end{tabular}\\
\end{center}
\end{table}

\section{Conclusion and Future Directions}
\label{sec:5}

In this paper, we introduced the MIMII dataset, a real-world dataset for investigating the malfunctioning behavior of industrial machines.
We collected 26,092 sound segments of normal condition and 6,065 sound segments of anomalous condition 
and mixed the background noise recorded in multiple real factories with the machine-sound segments for simulating real environments.
In addition, using the MIMII dataset, we presented our evaluation for autoencoder-based unsupervised anomalous sound detection.
We observed that non-stationary machine sound signals and noise are the key issues to be overcome in the development of an unsupervised anomaly detector. 
These results can be taken as a benchmark to improve the accuracy of anomaly detection in the MIMII dataset.

The MIMII dataset is freely available for download at \url{https://zenodo.org/record/3384388}.
To the best of our knowledge, this dataset is the first of its kind to address the problem of detecting anomalous conditions in industrial machinery through machine sounds. 
As benchmarking is an important aspect in data-driven methods, we believe that our MIMII dataset will be very useful to the research community. 
We are releasing this data to accelerate research in the area of audio event detection, specifically for machine sounds. 
This dataset can be applied to other use cases as well: for example, to restrict the training on a specific number of machine models and then test on the remaining machine models. 
This study will be useful for measuring the domain adaptation capability of the different methods applied on machines from different manufacturers. 
If the community takes an interest in our dataset and validates its usage, we will improve the current version with additional meta-data related to different anomalies. 

\bibliographystyle{IEEEtran}
\bibliography{refs}

\end{sloppy}
\end{document}